\documentclass[aps,pra,twocolumn,longbibliography]{revtex4-1}
\usepackage[normalem]{ulem}
\usepackage{float}
\usepackage{graphicx}  
\usepackage{dcolumn}          
\usepackage{amssymb}
\usepackage{appendix}
\usepackage{physics}   
\usepackage{mathtools}
\usepackage{esvect}
\usepackage{wrapfig}
\usepackage{amsthm}
\usepackage{verbatim}
\usepackage{bbm}
\usepackage[colorlinks=true,linkcolor=blue,citecolor=red,plainpages=false,pdfpagelabels]{hyperref}
\usepackage{xcolor}

\usepackage[mathscr]{euscript}
\usepackage{enumitem}

\usepackage{algorithm}
\usepackage[noend]{algpseudocode}
\usepackage{setspace}
\algblock[Class]{Class}{End}

\def\Tr{\operatorname{Tr}}

\def\({\left(}
\def\){\right)}
\def\[{\left[}
\def\]{\right]}

\newtheorem{corollary}{Corollary}

\newtheorem{observation}{Observation}

\newtheorem{proposition}{Proposition}

\def\>{\rangle}
\def\<{\langle}

\renewcommand{\arraystretch}{2}

\algnewcommand{\Inputs}[1]{
  \State \textbf{Inputs:}
  \Statex \hspace*{\algorithmicindent}\parbox[t]{.8\linewidth}{\raggedright #1}
}
\algnewcommand{\Initialize}[1]{
  \State \textbf{Initialize:}
  \Statex \hspace*{\algorithmicindent}\parbox[t]{.8\linewidth}{\raggedright #1}
}

\begin{document}
\renewcommand{\arraystretch}{1.5}
\widetext

\title{Testing of quantum nonlocal correlations under constrained free will and imperfect detectors}

\author{Abhishek Sadhu}
\email{abhisheks@rri.res.in}
\affiliation{Light and Matter Physics Group, Raman Research Institute, Bengaluru, Karnataka 560080, India}
\author{Siddhartha Das}
\email{das.seed@iiit.ac.in}
\affiliation{Center for Security, Theory and Algorithmic Research, Centre of Quantum Science and Technology, International Institute of Information Technology, Hyderabad, Gachibowli, Telangana 500032, India}

\date{\today}
\begin{abstract}
In this work, we deal with the relaxation of two central assumptions in standard locally realistic hidden variable (LRHV) inequalities: free will in choosing measurement settings, and the presence of perfect detectors at the measurement devices. Quantum correlations violating LRHV inequalities are called quantum nonlocal correlations. In principle, in an adversarial situation, there could be a hidden variable introducing bias in the selection of measurement settings, but observers with no access to that hidden variable could be unaware of the bias. In practice, however, detectors do not have perfect efficiency. A main focus of this paper is the introduction of the framework in which given a quantum state with nonlocal behavior under constrained free will, we can determine the threshold values of detector parameters (detector inefficiency and dark counts) such that the detectors are robust enough to certify nonlocality. We also introduce a class of LRHV inequalities with constrained free will, and we discuss their implications in the testing of quantum nonlocal correlations.
\end{abstract}

\maketitle

\section{Introduction}
The randomness acts as a primitive resource for various desirable information processing and computational tasks, e.g., cryptography~\cite{shannon1945mathematical,bennett1984proceedings,carl2006denial,Das18,das2021universal,primaatmaja2022security}, statistical sampling~\cite{yates1946review, dix2002chance}, probabilistic algorithms~\cite{rabin1980probabilistic,leon1988probabilistic,sipser1983complexity}, genetic algorithms~\cite{holland1992adaptation,morris1998automated,deaven1995molecular}, simulated annealing algorithms~\cite{ram1996parallel,fleischer1995simulated,kirkpatrick1983optimization}, neural networks~\cite{bishop1994neural,harrald1997evolving,zhang2000neural}, etc. Generation and certification of private randomness also form major goal of cryptographic protocols~\cite{acin2012randomness,colbeck2012free,RNGColbeck_2011,Das18,das2021universal,primaatmaja2022security}. The quantum theory allows for the generation and certification of true randomness~\cite{bell1964einstein,PhysRevLett.67.661,colbeck2012free,das2021universal,primaatmaja2022security}, which otherwise seems to be impossible within the classical theory~\cite{einstein1935can}.  

In a seminal paper~\cite{bell1964einstein}, J.S.~Bell showed that the statistical predictions of the quantum mechanics cannot be explained by local realistic hidden variable (LRHV) theories. Several LRHV inequalities, also called Bell-type inequalities, that are based on two physical assumptions--- existence of local realism and no-signalling criterion~\cite{einstein1935can}, have been derived since then (see e.g., ~\cite{brunner2014bell,HSD15} and references therein). Quantum systems violating these LRHV inequalities~\cite{brunner2014bell} are said to have quantum nonlocal correlations. Quantum nonlocal correlations are not explainable by any LRHV theories. Hence, the randomness generated from the behavior of quantum systems with nonlocal correlations can be deemed to exhibit true randomness. Presence of these nonlocal correlations allow for the generation and certification of randomness and secret key in a device-independent way~\cite{PhysRevLett.67.661,QKDAcin2006,acin2016certified,kaur2021upper, primaatmaja2022security,zapatero2022advances,wooltorton2022tight}.

It is known that the experimental verification of the Bell's inequality requires additional assumptions which could lead to incurring loopholes such as locality loophole~\cite{bell2004speakable}, freedom-of-choice (measurement independence or free will) loophole~\cite{clauser1974experimental,bell1985exchange}, fair-sampling loophole (detection loophole)~\cite{pearle1970hidden,clauser1974experimental}. In recent major breakthroughs~\cite{hensen2015loophole,giustina2015significant,shalm2015strong,big2018challenging}, incompatibility of quantum mechanics with LRHV theories has been demonstrated by considerably loophole-free experiments showing violation of Bell's inequality by quantum states with quantum nonlocal correlations. In an experimental setup, locality assumption requires space like separation between the measurement events~\cite{einstein1935can,bell1964einstein} which leads to locality loophole~\cite{aspect1982experimental,weihs1998violation}. In a Bell experiment with photonic setup, the detection events of the two parties are identified as belonging to the same pair, by observing whether the difference in the time of detection is small~\cite{larsson2014bell}. It makes possible for an adversary to fake the observed quantum correlations by exploiting the time difference between the detection events of the parties~\cite{larsson2004bell} giving rise to the detection loophole, which can be closed by using predefined coincidence detection windows~\cite{larsson2004bell}.

The free will assumption~\footnote{It should be noted that the free will assumption mentioned in this paper is also called measurement independence. This assumption relates to the possible correlations between the choice of measurement settings for the two parties, which can affect the observed experimental statistics.} in the Bell-type inequality states that the parties (users) can choose the measurement settings freely or, use uncorrelated random number generators. In an experimental demonstration presented in~\cite{big2018challenging} human random decisions were used to choose the measurement settings in an attempt to close this loophole. In practice, the detectors are imperfect which introduce dark counts~\cite{eisaman2011invited} and detection inefficiency\cite{eisaman2011invited}. In this paper, we discuss implications of bias in the freedom-of-choice and non-ideal detectors on the LRHV tests of quantum nonlocal correlations.
\begin{figure}
    \includegraphics[scale=0.6]{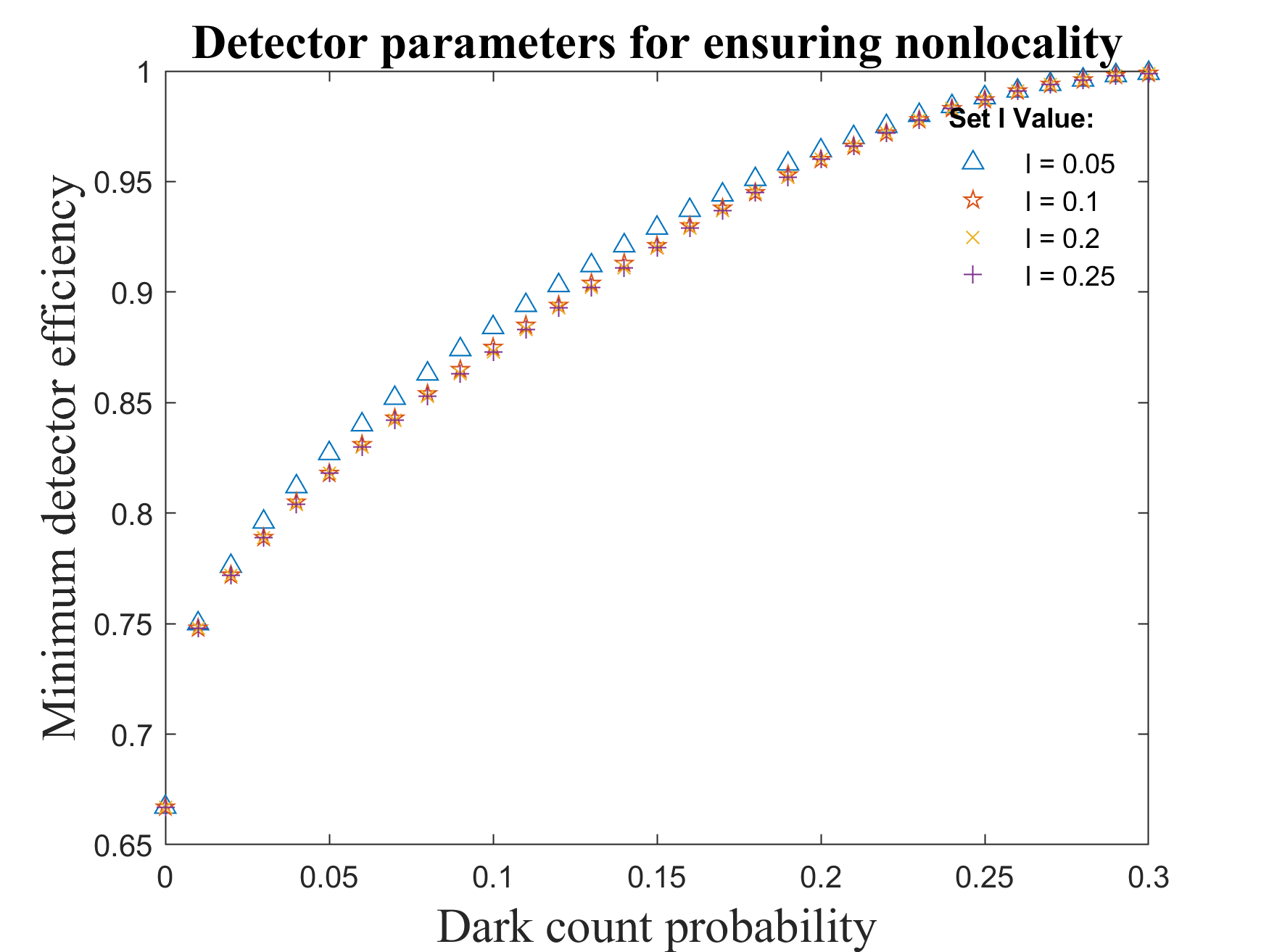}
    \caption{In this figure we plot the minimum detection efficiency $\eta$ as a function of the dark count probability $\delta$ obtained using Algorithm~\ref{algo:Gisin}. For each pair of $(\eta,\delta)$ in this figure, there exists a quantum behavior $\{p(a^{id},b^{id}|xy)\}$ that violates Eq.~\eqref{MDLeq12}. From such a behavior, $\{p(a^{ob},b^{ob}|xy)\}$ is obtained by using the values of the pair $(\eta,\delta)$. The behavior $\{p(a^{ob},b^{ob}|xy)\}$ violates Eq.~\eqref{SauerInequality}. See Section~\ref{sec:ImperfectDet}. (Color Online)}
    \label{fig:critParGisin}
\end{figure}
To the best of our knowledge, the assumption of free will was relaxed in \cite{hall2010local} using a distance measure based quantification of the measurement dependence. It was shown that the Bell-CHSH inequality can be violated by sacrificing equal amount of free will for both parties. This result was extended to the scenario of parties having different amounts of free will in~\cite{friedman2019relaxed} and for one of the parties in~\cite{banik2012optimal}. In an alternate approach, measurement dependence was quantified in~\cite{putz2014arbitrarily,putz2016measurement} by bounding the probability of choosing of the measurement settings to be in a given range. Following this approach, tests for nonlocality has been constructed \cite{putz2014arbitrarily,zhao2022tilted}. These inequalities has been applied to randomness amplification protocols~\cite{kessler2020device,zhao2022tilted}. However, the consideration of imperfect detectors in the implication of these measurement-dependent LRHV inequalities are still lacking as the above mentioned works assumed perfect detection while allowing for relaxation in measurement dependence.  

A main focus of this work is to study the implications of imperfect detectors and constrained free will on the test of quantum nonlocal correlations. We adapt the approach discussed in \cite{sauer2020quantum} to model imperfect detectors for the Bell experiment as a sequential application of a perfectly working inner box followed by a lossy outer box. The inner box contains a quantum source whose behavior is nonlocal under constrained free will, i.e., violates certain measurement dependent LRHV inequality. The outer box separately introduces detector inefficiency and dark counts for each party. Using this model, we determine the threshold values of the detector parameters that make detectors robust for testing of quantum nonlocality under constrained free will (e.g.,~see Fig.~\ref{fig:critParGisin} with details in Section~\ref{sec:ImperfectDet}). Next for the scenario of perfect detectors, we compare the implications two different approaches presented in~\cite{putz2014arbitrarily} and~\cite{hall2010local} to quantify measurement dependence (a) by bounding the probability of choosing the measurement settings $x$ (for Alice's side) and $y$ (for Bob's side) conditioned on a hidden variable $\lambda$ to be in the range $[l,1-3l]$~\cite{putz2014arbitrarily} and (b) by using a distance measure $M$ to quantify measurement settings distinguishability~\cite{hall2010local}. This comparison is made in the 2 (party) - 2 (measurement settings per party) - 2 (outcome per measurement) scenario and their effects on the certification of the nonlocality. We also introduce a new set of measurement dependent LRHV (MDL) inequalities by introducing distance-based measurement dependent quantity in adapted AMP tilted Bell inequality\cite{acin2012randomness} and discuss implications and trade-off between measurement dependence parameters and tilted parameters for the certification of quantum nonlocal correlations.

Throughout this paper we limit our discussions to LRHV for bipartite quantum systems. The structure of this paper is as follows. In Section~\ref{AdverserySettings}, we introduce the framework of locality and measurement dependence, which is a constraint that limits the  free will of the user. We present a model where the user is tricked by an adversary into thinking that they have freedom-of-choice for the measurement. In Section~\ref{MDLBounding}, we consider and compare two different approaches to quantify the measurement dependence in terms of parameters $l$ and $M$ mentioned earlier. We determine the critical values of these parameters necessary for the certification of nonlocality in the measurement dependence settings and compare them with amount of violation obtained for the Bell-CHSH inequality, tilted Hardy relations and the tilted Bell inequalities. In Section~\ref{sec:ImperfectDet}, We determine the threshold values of the detector parameters namely inefficiency and dark count probability such that the detectors are robust enough to certify nonlocality in presence of constrained free will. In Section~\ref{sec:TiltedBell}, we introduce a new set of LRHV inequalities adapted from the AMP tilted Bell inequality using distance measures based measurement dependence quantities. We use these inequalities to observe the effect of relaxing the free will assumption for either parties on the certification of quantum nonlocal correlations.

\section{The adverserial role in choice of measurement settings} \label{AdverserySettings}
Formally a quantum behavior of a device with bipartite state $\rho_{AB}$ is given by its quantum representation $\{p(ab|xy)\}$, where $p(ab|xy)=\Tr[\Gamma^{x}_a\otimes \Gamma^{y}_b \rho_{AB}]$ with $\{\Gamma^{x}_a\}_a$ and $\{\Gamma^{y}_b\}_b$ being Positive Operator Valued Measures (POVMs) for each input $(x,y)$ to the device, for $x\in\mathcal{X}$ and $y\in\mathcal{Y}$. If a state $\rho_{AB}=\op{\psi}_{AB}$ is pure, then we may simply represent it as a ket $\ket{\psi}_{AB}$ instead of density operator $\rho_{AB}$. Henceforth, the abbreviation MDL inequalities would stand for measurement dependent LRHV inequalities.

We consider the Bell scenario where two parties, Alice and Bob, share a bipartite quantum state $\rho_{AB}$. Each party can choose to perform one of two measurements available to them, i.e., $|\mathcal{X}|=2=|\mathcal{Y}|$. We attribute POVM $\{\Lambda^x_a\}_x$ to Alice and $\{\Lambda^y_b\}_y$ to Bob. Each of these measurements can have two outcomes. We denote measurement outcomes for Alice and Bob by $a$ and $b$, respectively. The statistics of the measurement outcomes in the experiment can then be described by the probability distribution $\textbf{P} = \{p(ab|xy)\}$ which is also termed as behavior. In this framework, let there exist a hidden variable $\lambda$ belonging to some hidden-variable space, $\Lambda$. The probability distribution of the outputs conditioned on the inputs can then be expressed as
\begin{equation} \label{LHHVdef}
    p(ab|xy) = \sum_{\lambda\in\Lambda} p(ab|xy\lambda) p(\lambda|xy).
\end{equation}
The hidden variable $\lambda$ (distribution according to $p(\lambda)$) can provide an explanation of the observed experimental (measurement) statistics. In each run of the experiment, there exists a fixed $\lambda$ that describes the outcome of the experimental trial following the distribution $p(ab|xy\lambda)$. After multiple runs of the experiment, the output statistics are described by sampling from the distribution $p(\lambda|xy)$. 

In an adversarial scenario, Alice and Bob can believe that they are choosing all the settings with equal probability, i.e., $p(xy) = 1/4$ for each pair $(x,y)$ while an adversary biases their choice in the scale $\lambda$. The adversary can distribute the settings chosen by Alice and Bob according to
\begin{equation} \label{eq:AdversarySettingDist}
    p(xy) = \sum_{\lambda\in\Lambda} p(xy|\lambda) p(\lambda).
\end{equation}
In Eq.~\eqref{eq:AdversarySettingDist}, let $\lambda$ take two values, $\lambda_1$ and $\lambda_2$, whose probability distributions are given as $p\text{$(\lambda_1 $)}=\sin ^2(\theta_\lambda )$ and $ p\text{$(\lambda_2 $)}=\cos ^2(\theta_\lambda )$, respectively. In the simplest scenario $x$ and $y$ can each take values $0$ or $1$ and there are four possible ways in which the measurement settings can be chosen by Alice and Bob, i.e., $(x,y) \in \{(0,0), (0,1), (1,0), (1,1)\}$. Let the probability of choosing the measurement setting $(x,y)$ conditioned on the hidden variable be distributed according to Table~\ref{Table1}.
\begin{table}[t]
    \begin{tabular}{|p{2cm}||p{3cm}||p{3cm}|}
    \hline
    \multicolumn{3}{|c|}{Probability distributions of Eve} \\
    \hline
    Joint Setting & Distribution 1 $(\lambda_1)$ & Distribution 2 $(\lambda_2)$ \\
     \hline
    $p(0,0|\lambda)$ & $\cos ^2(\phi_{s_1})$ & $\cos ^2(\phi_{s_2})$  \\
     \hline
    $p(0,1|\lambda)$ & $\sin ^2(\theta_{s_1}) \sin ^2(\phi_{s_1})$  & $\sin ^2(\theta_{s_2}) \sin ^2(\phi_{s_2})$ \\
     \hline
    $p(1,0|\lambda)$ & $\cos^2(\delta_{s_1}) \cos^2(\theta_{s_1}) \newline \sin^2(\phi_{s_1})$ & $\cos ^2(\delta_{s_2}) \cos^2(\theta_{s_2}) \newline \sin^2(\phi_{s_2})$ \\
     \hline
    $p(1,1|\lambda)$ & $\sin^2(\delta_{s_1}) \cos^2(\theta_{s_1}) \newline \sin ^2(\phi_{s_1})$  & $\sin ^2(\delta_{s_2}) \cos^2(\theta_{s_2}) \newline \sin^2(\phi_{s_2})$\\
     \hline
    \end{tabular}
    \caption{Probability distribution for choice of settings by Alice and Bob based on the hidden variable $\lambda$. An adversary can use this distribution and by suitably choosing the parameters of the table trick Alice and Bob into thinking they have free will in choosing the measurement settings.}
    \label{Table1}
\end{table}
To observe the effect of the conditional probability distribution for choice of $x$ and $y$ from Table~\ref{Table1} on $p(xy)$, consider the following examples: 
\begin{enumerate}
    \item[(i)] For the case of $\theta_{s_2}=0.6847$, $\phi_{s_2}=1.2491$, $\phi_{s_1}=0.8861$, $\delta_{s_2}=1.2491$, $\theta_{\lambda}=0.785398$, $\theta_{s_1}=0.50413$, and $\delta_{s_1}=0.175353$. It can be seen that this choice of parameters set $p(xy) = 0.25~\forall~(x,y)$.
    \item[(ii)] For the case of $\theta_{s_2}=0.5796$, $\phi_{s_2}=1.2491$, $\phi_{s_1}=0.6847$, $\delta_{s_2}=0.75$, $\theta_{\lambda} = 0.57964$, $\theta_{s_1}=0.793732$, and $\delta_{s_1}=1.06395$. It can be seen that this choice of parameters set $p(xy) = 0.25~\forall~(x,y)$.
\end{enumerate}
The above examples show that by proper choice of parameters in Table~\ref{Table1}, the adversary can trick Alice and Bob into thinking that they have free will in choosing the measurement setting. In the scenario where Alice and Bob chooses between the measurement settings with unequal probabilities, i.e., $p(xy) \neq 1/4$ for each pair (x,y), the adversary can adjust the parameters of Table~\ref{Table1} accordingly. In the presence of a bias in the choice of measurement settings in the $\lambda$ scale which Alice and Bob are unaware of, the following constraints \cite{blasiak2021violations} can be imposed on the conditional joint probability distribution:
\begin{itemize}
    \item[a.] The signal locality, i.e., no-signalling, assumption imposes the factorisability constraint on the conditional joint probability distribution,
\begin{equation} \label{locality}
    p(ab|xy\lambda) = p(a|x\lambda) p(b|y\lambda).
\end{equation}
    \item[b.] The measurement independence, i.e., freedom-of-choice or free will, assumption requires that $\lambda$ does not contain any information about $x$ and $y$ which is equivalent to stating
\begin{eqnarray}
&& p(\lambda|xy) = p(\lambda) \nonumber \\ 
\text{or equivalently, } \hspace{5pt} && p(xy|\lambda) = p(xy). \label{measerementInDep}
\end{eqnarray}
\end{itemize}

\section{Quantifying measurement dependence} \label{MDLBounding}
In this section, we first review two different approaches considered in ~\cite{putz2014arbitrarily,zhao2022tilted} and  \cite{hall2010local,banik2012optimal,friedman2019relaxed} to quantify the measurement dependence. We then compare these two approaches and observe their effects on the tests of quantum nonlocal behaviors.   

\subsection{Review of MDL inequalities from prior works} \label{reviewQuantification}
We review approach discussed in~\cite{putz2014arbitrarily,putz2016measurement} to quantify measurement dependence by bounding the probability of choice of measurement settings conditioned on a hidden variable to be in a specific range (Section~\ref{rev1}). Then we review approach discussed in \cite{hall2010local,banik2012optimal,friedman2019relaxed} to quantify measurement dependence using a distance measure (Section~\ref{sec:QuantificationApproach2}).   
\subsubsection{Bound on the probability of choosing the measurement settings} \label{rev1}
In the works \cite{putz2014arbitrarily,putz2016measurement}, we observe that the probability of Alice and Bob to choose measurement settings $x$ and $y$ conditioned on $\lambda$ can be bounded as 
\begin{equation} \label{eqn:MDLpardef1}
    l \leq p(xy|\lambda)\leq h,
\end{equation}
where $0 \leq l \leq p(xy|\lambda) \leq h \leq 1$. If Alice and Bob each chooses from two possible values of measurement settings, then $l = h = 0.25$ corresponds to the complete measurement independence; other values of $l$ and $h$ represent bias in the choice of measurement settings.

\begin{figure}
    \includegraphics[scale=0.35]{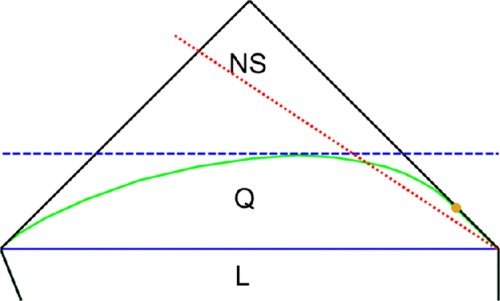}
    \caption{A two dimensional slice of the no-signalling space with the MDL correlations discussed in \cite{putz2014arbitrarily}. The blue line encloses the set of Bell-CHSH local correlations. The green line encloses the quantum set. The black triangle encloses the no-signalling distributions. For the case of $h = 1 - 3l$, the inequality ~\eqref{temp} shifts from the Bell-CHSH boundary to the no-signalling boundary via the red dotted line. (Color online)}
    \label{fig:MDLslice}
\end{figure}

In the 2 (user) - 2 (measurement settings per user) - 2 (outcome per measurement) scenario with $a,b \in\{+,-\}$ and $x,y \in\{0,1\}$, it was shown in \cite{putz2014arbitrarily, putz2016measurement} that all the measurement dependent local correlations satisfy the PRBLG MDL inequality 
\begin{equation} \label{temp}
    l  p(++00) - h (p(+-01)+p(-+10)+p(++11)) \leq 0.
\end{equation}
A two-dimensional slice in the non-signalling space is shown in Fig.~\ref{fig:MDLslice} (figure from \cite{putz2014arbitrarily}). In the figure, the quantum set is bounded by the green line and the set of non-signalling correlations lie within the black triangle.
In Fig.~\ref{fig:MDLslice}, the red dotted line corresponds to Eq.~\eqref{temp} with $h = 1 - 3l$. If we set $h = 1 - 3l$, the PRBLG MDL inequality tilts from the Bell-CHSH inequality ($l = 0.25$) to the non-signalling border ($l = 0$). For $h = 1 - 3l$, Eq.~\eqref{temp} is expressed as
\begin{eqnarray} 
&& l p(++00) - (1 - 3l) (p(+-01)+p(-+10) \nonumber \\
&& +p(++11)) \leq 0. \label{MDLeq1}
\end{eqnarray}
We note that if Alice and Bob believes they have complete measurement dependence, i.e., $p(xy) = 0.25 \hspace{3pt} \forall (x,y)$, then Eq.~\eqref{MDLeq1} reduces to 
\begin{eqnarray}
&& l p(++|00) - (1 - 3l) (p(+-|01) \nonumber \\
    &&+p(-+|10)+p(++|11)) \leq 0. \label{MDLeq12}
\end{eqnarray}
It follows from \cite{zhao2022tilted} that invoking the PRBLG MDL inequality~\eqref{MDLeq1} in the tilted Hardy test~\cite{zhao2022tilted}, we obtain the ZRLH MDL inequality. The ZRLH MDL inequality is expressed as
\begin{eqnarray}
&& l [p(++|00) + w p(--|00) - \max\{0,w\}] \nonumber \\
&& -(1 - 3l) [p(+-|01) + p(-+|10) + p(++|11)] \leq 0, \nonumber \\ \label{RamanathanMDL} 
\end{eqnarray} 
where $w$ is the tilting parameter taking real numbers in the range, $w \in (-0.25,1)$. We call the quantum behaviors that violate Eqs.~\eqref{MDLeq12}~and~\eqref{RamanathanMDL} as quantum nonlocal in the presence of measurement dependence.

\subsubsection{Distance measure to quantify measurement distinguishability} \label{sec:QuantificationApproach2}
We discussed in Section~\ref{AdverserySettings} that the experimental statistics described by the joint probability distribution $p(ab|xy)$ can be explained by $\lambda \in \Lambda$ in the following form,
\begin{equation}
    p(ab|xy) = \int {\rm d}\lambda p(ab|xy\lambda) p(\lambda|xy). 
\end{equation}
The assumption of the measurement independence constrains the probability distribution of measurement settings via
\begin{equation}
    p(\lambda|xy) = p(\lambda). \label{eq:B2}
\end{equation}
Eq.~\eqref{eq:B2} implies that no extra information about $\lambda$ can be obtained from the knowledge of $x$ and $y$. This is equivalent to saying
\begin{equation}
    p(xy|\lambda) = \frac{p(\lambda|xy) p(xy)}{p(\lambda)} = p(xy). \label{eq:B3}
\end{equation}
Eq.~\eqref{eq:B3} implies that Alice and Bob has complete freedom in choosing the measurement settings $x$ and $y$ respectively. If $x\in U\equiv\{x_1,x_2\}$ and $y\in V\equiv\{y_1,y_2\}$, measurement dependence implies $p(\lambda|x_1,y_1) \neq p(\lambda|x_2,y_2)$. Distinguishability between $p(\lambda|x_1,y_1)$ and $p(\lambda|x_2,y_2)$ can be quantified using a distance measure defined in~\cite{hall2010local},
\begin{equation} \label{quantify}
    M = \int {\rm d}\lambda |p(\lambda|x_1,y_1) - p(\lambda|x_2,y_2)|.
\end{equation}
We can express the probability to successfully distinguish $p(\lambda|x_1,y_1)$ and $p(\lambda|x_2,y_2)$ based on the knowledge of $\lambda$ as
\begin{equation} \label{F}
    F = \frac{1}{2} \left(1 + \frac{M}{2}\right).
\end{equation}
When $M = 0$, we have $F = \frac{1}{2}$, which suggests that no additional information about the hidden variable $\lambda$ can be obtained from knowing the choice of measurement settings.

This observation is consistent with maximum free will that Alice and Bob has while choosing the measurement settings. Whereas for $M = 2$, we have $F = 1$, which suggests that the complete information about the hidden variable $\lambda$ can be obtained from knowing the choice of the measurement settings. This observation is consistent with no free will for Alice and Bob while choosing the measurement settings.

The local degrees of measurement dependence for Alice and Bob as introduced in~\cite{hall2010local} is given by
\begin{widetext}
\begin{equation} \label{M1}
    M_1 \equiv \max \left\{ \int {\rm d}\lambda \abs{p(\lambda|x_1,y_1) - p(\lambda|x_2,y_1)}, \int {\rm d}\lambda \abs{p(\lambda|x_1,y_2) - p(\lambda|x_2,y_2)}\right\},
\end{equation}
\begin{equation} \label{M2}
    M_2 \equiv \max \left\{ \int {\rm d}\lambda \abs{p(\lambda|x_1,y_1) - p(\lambda|x_1,y_2)}, \int {\rm d}\lambda  \abs{p(\lambda|x_2,y_1) - p(\lambda|x_2,y_2)}\right\},
\end{equation}
\end{widetext}
$M_1$ quantifies the measurement dependence for Alice's settings keeping Bob's settings fixed, and similarly the other way round for $M_2$. 
The above parameters will be useful in deriving the bounds on the AMP tilted Bell inequalities \cite{acin2012randomness} in the Section~\ref{sec:TiltedBell}.

\subsection{Comparison and discussion on implications of different measurement dependent LRHV inequalities}
In this section, we check for the allowed values of measurement dependence parameter $l$ to ensure nonlocality of quantum behaviors that are used to obtain randomness. 
\begin{table}
    \centering
    \begin{tabular}{p{3.8cm}  p{4.5cm}}
    \hline
    \hline
    \multicolumn{2}{c}{Expectation values of the operators} \\
    \hline
    \hline
    $\langle x_1 \rangle = \cos (2 \phi );$ & $\langle x_2 \rangle = 0;$ \\
    $\langle y_1 \rangle = \cos (2 \phi ) \cos (\mu );$ & $\langle y_2 \rangle = \cos (2 \phi ) \cos (\mu );$ \\
    $\langle x_1,y_1 \rangle = \cos (\mu );$ & $\langle x_1,y_2 \rangle = \cos (\mu );$ \\
    $\langle x_2,y_1 \rangle = \sin (2 \phi ) \sin (\mu );$ & $\langle x_2,y_2 \rangle = -\sin (2 \phi ) \sin (\mu )$. \\
     \hline
    \end{tabular}
    \caption{The table presents the expectation values of operators from \cite{acin2012randomness} that violate the AMP tilted Bell inequality given in Eq.~\eqref{TBI}. The parameter $\mu := \tan^{-1} (\sin (2\phi) / \alpha)$ where $\phi$ is the state parameter and $\alpha$ is the tilting parameter in Eq.~\eqref{TBI} as defined in \cite{acin2012randomness}.}
    \label{table:AMPTltedBellMaxViolation}
\end{table}
Consider the quantum behavior in Table \ref{table:AMPTltedBellMaxViolation} that violates the AMP tilted Bell inequality~\eqref{TBI}. In the limit of $\alpha \rightarrow \infty$, close to two bits of randomness can be obtained from such a behavior by violating the AMP tilted Bell inequality~\cite{acin2012randomness}. For the quantum behavior in Table \ref{table:AMPTltedBellMaxViolation}, the PRBLG MDL inequality~\eqref{MDLeq12} reduces to
\begin{equation} \label{MDLTb}
l \leq \frac{3 \alpha  t +\alpha  \cos (2 \phi ) t-2 \sqrt{2} \alpha  \sin ^2(\phi )+\sqrt{2} (\cos (4 \phi )-1)}{10 \alpha  t+4 \alpha  \cos (2 \phi ) \left(t+\sqrt{2}\right)-2 \sqrt{2} \left(\alpha +3 \sin ^2(2 \phi )\right)}
\end{equation}
where $t = \sqrt{-\frac{\cos (4 \phi )}{\alpha ^2}+\frac{1}{\alpha ^2}+2}$. In the limit of $\alpha \rightarrow \infty$, from Eq.~\eqref{MDLTb} we have $l \leq 0.25$. 
We see that in the limit of $\alpha \rightarrow \infty$ the quantum behavior in Table~\ref{table:AMPTltedBellMaxViolation} does not violate any PRBLG MDL inequality. For $\alpha = 1$ (which is equivalent to Bell-CHSH inequality) and $\phi = \frac{\pi}{4}$ (the maximum violation of the Bell-CHSH inequality,) we see from Eq.~\eqref{MDLTb} that the quantum behavior in Table~\ref{table:AMPTltedBellMaxViolation} violates the family of PRBLG MDL inequalities for $l > 0.2023$.

\begin{figure}
    \includegraphics[scale = 0.6]{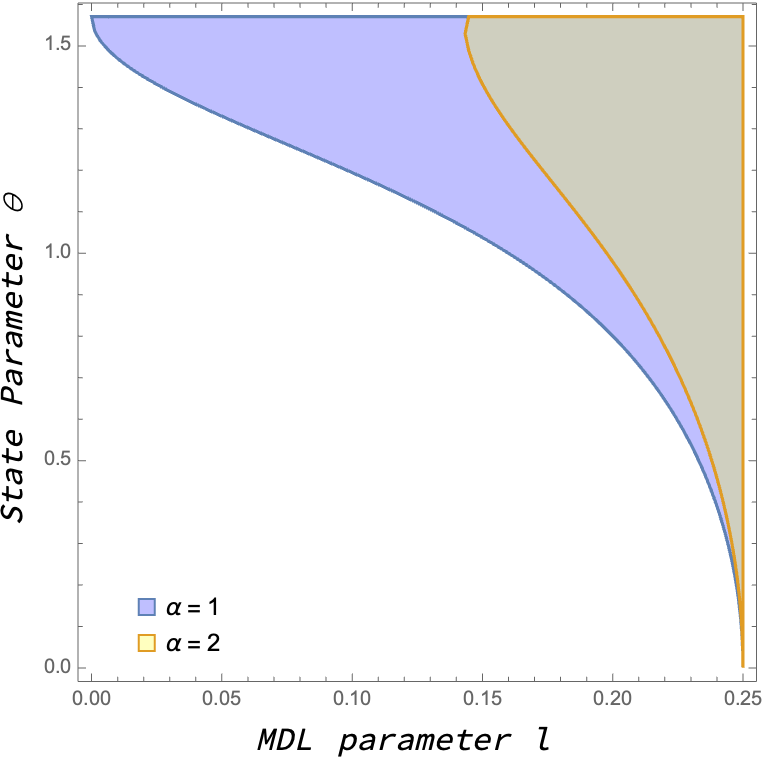}
    \caption{For $\alpha \in \{1,2\}$, we plot the range of the state parameter $\phi$ from Table~\ref{table:AMPTltedBellMaxViolation} $(\phi=\frac{1}{2}\sin^{-1}(\alpha \tan \mu))$ that violates the PRBLG MDL inequality (given by Eq.~\eqref{MDLeq12}) for different values of the measurement dependence parameter $l$. (Color online)}
    \label{fig:my_label}
\end{figure}

We observe in Fig.~\ref{fig:my_label} that for a fixed value of $\alpha$, the range of the allowed values of $\phi$ from Table~\ref{table:AMPTltedBellMaxViolation} that violates the PRBLG MDL inequality given by Eq.~\eqref{MDLeq12} increases with the increase in $l$. Also as $\alpha$ increases, for a particular value of $l$, there is a decrease in the range of the allowed values of $\phi$ from Table~\ref{table:AMPTltedBellMaxViolation} that violates the AMP tilted Bell inequality. It was shown in~\cite{zhao2022tilted} that the state $\rho_{g}=\op{\psi_g}$~\eqref{eq:psi-g} and measurement settings $\{A_0^g, A_1^g, B_0^g, B_1^g\}$ can be used to obtain close to $1.6806$ bits of global randomness (at $\theta \approx 1.13557$), where
\begin{equation}\label{eq:psi-g}
\ket{\psi_g} = \cos(\frac{\theta}{2}) \ket{00} - \sin(\frac{\theta}{2}) \ket{11},    
\end{equation}
\begin{eqnarray}
 A_0^g = B_0^g = && \frac{ -\sqrt{2} \sin (\theta ) \sqrt{\sin (\theta )}}{(2-\sin (\theta )) \sqrt{\sin (\theta )+1}} \sigma_x \nonumber\\ 
 && + \frac{ -(\sin (\theta )+2) \sqrt{1-\sin (\theta )}}{(2-\sin (\theta )) \sqrt{\sin (\theta )+1}} \sigma_z,
\end{eqnarray}
\begin{eqnarray}
A_1^g = B_1^g = \frac{ \sqrt{2} \sqrt{\sin (\theta )}}{\sqrt{\sin (\theta )+1}} \sigma_x  -\frac{ \sqrt{1-\sin (\theta )}}{\sqrt{\sin (\theta )+1}} \sigma_z,
\end{eqnarray}
such that $\theta = \sin^{-1}{(3-\sqrt{4w + 5})}$ and $A_x^g,B_y^g$ for $x,y\in\{0,1\}$ denote measurement settings for Alice and Bob, respectively. 
\begin{observation} \label{obs:4}
Let us consider the quantum behavior given by the state $\rho_g$ and measurement settings $\{A_0^g, A_1^g, B_0^g, B_1^g\}$. For such a behavior, the PRBLG MDL inequality given by Eq.~\eqref{MDLeq12} reduces to
\begin{align}
& \frac{2l}{(\xi-1)^2}\left(31 \xi+w \left(4 \xi-34\right)-69\right) \nonumber\\
& + \frac{2l(\xi-5)\sqrt{\xi-2}}{\sqrt{4-\xi}(\xi-1)}\sqrt{-4 w+6 \xi-13}\nonumber\\
& - \frac{2(1-3l)}{1-\xi}\bigg( 8w-10\xi+\frac{\sqrt{\xi-2} \sin \left(2 \sin ^{-1}\left(3-\xi\right)\right)}{\sqrt{4-\xi}} \nonumber\\ 
&   + 22 \bigg) \leq 0, \label{obs:21}
\end{align}
where $\xi = \sqrt{4 w+5}$. On inspection we see that for all $w \in (-0.25,1)$, Eq.~\eqref{obs:21} reduces to $l \leq 0$. The quantum behavior specified by the state $\rho_g$ and measurement settings $\{A_0^g, A_1^g, B_0^g, B_1^g\}$ is nonlocal for $l > 0$.

For the given quantum behavior, the ZRLH MDL inequality~\eqref{RamanathanMDL} reduces to
\begin{eqnarray}
&& \frac{1}{2 \left(\xi-1\right)}(2 l \left(-4 w+7 \xi-15\right)+ \nonumber \\
    &&\frac{(1-3 l) \sqrt{\xi-2} \sin \left(2 \sin ^{-1}\left(3-\xi\right)\right)}{\sqrt{4-\xi}}+8 w-10 \xi+22) \nonumber \\
&&-l \max (0,w) \leq 0. \label{obs:22}
\end{eqnarray}
On inspection we see that for all $w \in (-0.25,1)$, Eq.~\eqref{obs:22} reduces to $l \leq 0$.

That is, the quantum behavior given by state $\rho_g$ and measurement settings $\{A_0^g, A_1^g, B_0^g, B_1^g\}$ violates both PRBLG and ZRLH MDL inequalities and its quantum nonlocality can be certified for all possible $l > 0$.
\end{observation}

\section{Imperfect detector and constrained free will} \label{sec:ImperfectDet}
We model the detection units of Alice and Bob using a two-box approach following~\cite{sauer2020quantum}. There is an inner box containing a quantum source generating bipartite quantum states whose behavior is nonlocal under constrained free will but assuming that detectors are perfect. Nonlocality of the quantum behavior in the inner box is tested by violation of a given MDL inequality. The output of the inner box is quantum nonlocal behavior that violates a given MDL inequality. An outer box introduces the detector imperfections, namely the detection inefficiency and dark counts. The quantum nonlocal behavior obtained from the inner box gets mapped at the outer box to the output behavior with detector imperfection parameters.  The output behavior then undergoes an LRHV test based on which we are able to determine threshold values of detection inefficiency and dark counts such that the given quantum nonlocal behavior can still be certified to be nonlocal with imperfect detectors. A deviation from a two-box approach in~\cite{sauer2020quantum} is the introduction of the measurement dependence assumption to the working of the inner box. Alice and Bob has access only to the input settings and the outputs of the outer box. We assume that either party (Alice and Bob) has access to two identical detectors that can distinguish between the orthogonal outputs. 

The measurement outcomes for the inner box are labeled as $a^{id}$ and $b^{id}$ respectively. We note that $a^{id}$ and $b^{id}$ can each take values from the set $\{+,-\}$. Introducing non-unit detection efficiency, $0\leq\eta\leq1$, and non-zero dark count probability, $0\leq\delta\leq1$ in the outer box, the ideal two outcome scenario becomes a 4-outcome scenario with the addition of no-detection event $\Phi$, and the dark-count event $\chi$. These events are defined in the following way:
\begin{itemize}
    \item[$\Phi:$] One particle is sent to the party and none of the two detectors of that party click.
    \item[$\chi:$] One particle is sent to the party and both the detectors of the party click.
\end{itemize}
We label the measurement outcomes for Alice and Bob obtained from the outer box as $a^{ob}$ and $b^{ob}$ respectively. We note that $a^{ob}$ and $b^{ob}$ can each take values from the set $\{+,-,\Phi,\chi\}$. Furthermore, we assume that Alice and Bob's detection units have identical values for $\eta$ and $\delta$. The conditional probability of observing the outcome $t^{ob}$ from the outer box conditioned on observing $t^{id}$ in the ideal scale is given by $p(t^{ob}|t^{id})$ with $t^{ob}\in \{a^{ob},b^{ob}\}$ and $t^{id}\in \{a^{id},b^{id}\}$. The observed joint probabilities can then be expressed as~\cite{sauer2020quantum}
\begin{equation} \label{eq:IdealToObserved}
    p(a^{ob}b^{ob}|xy) = \sum_{a^{id},b^{id}} p(a^{ob}|a^{id}) p(b^{ob}|b^{id})  p(a^{id} b^{id}|xy).
\end{equation}

We relax the free will assumption in the inner box by introducing the hidden variable $\lambda \in \Lambda$. Considering this assumption, $p(a^{id} b^{id}|xy)$ is expressed as
\begin{equation}
    p(a^{id} b^{id}|xy) = \sum_{\lambda\in\Lambda} p(a^{id} b^{id}|xy\lambda)  p(\lambda|xy).
\end{equation}
The hidden variable, $\lambda$ (distributed according to $p(\lambda)$) provides an explanation of the observed experimental statistics of the inner box. The distribution of settings that are chosen by Alice and Bob depends on $\lambda$ via the following relation,
\begin{equation}
    p(xy) = \sum_{\lambda\in\Lambda} p(xy|\lambda) p(\lambda).
\end{equation}
If we impose the locality condition from Eq.~\eqref{locality} on the experimental statistics of the inner box, we arrive at the following factorisability constraint,
\begin{equation}
    p(a^{id} b^{id}|xy\lambda) = p(a^{id}|x\lambda) p(b^{id}|y\lambda).
\end{equation}
Also, if we impose the measurement independence assumption from Eq.~\eqref{measerementInDep}, we arrive at the following constraint,
\begin{equation}
    p(xy|\lambda) = p(xy)~ \text{or equivalently,}~ p(\lambda|xy) = p(\lambda).
\end{equation}
The output statistics of the outer box for imperfect detectors can depend on the output of the inner box in the following four ways \cite{sauer2020quantum}:
\begin{itemize}
    \item[i)] No particle is detected on either of the detectors and no dark count detection event takes place. We can then write the following:
    \begin{equation}
        p(a^{ob}|a^{id}) = (1-\eta)(1-\delta)^2. \label{det:case1}
    \end{equation}
    
    \item[ii)] No particle is detected by the detector that should have detected it and a dark count takes place in the other detector. We can then write the following:
    \begin{equation}
        p(a^{ob}|a^{id}) = (1-\eta)(1-\delta)\delta. \label{det:case2}
    \end{equation}
    
    \item[iii)] Either the particle is detected by one of the detectors and a dark count takes place in the other detector or the particle is not detected and dark counts take place in both the detectors. We can then write the following:
    \begin{eqnarray}
    p(a^{ob}|a^{id}) &&= \eta \delta + (1-\eta)\delta^2 \nonumber \\
    &&= \delta [1 - (1-\eta)(1-\delta)].    \label{det:case3}
    \end{eqnarray}
    
    \item[iv)] Either the particle is detected and no dark count takes place or the particle is not detected and a dark count takes place in the detector in which the particle should have been registered. We can then write the following:
    \begin{eqnarray}
    p(a^{ob}|a^{id}) &&= \eta (1-\delta) + (1-\eta)\delta (1-\delta) \nonumber \\
    &&= (1-\delta) [1 - (1-\eta)(1-\delta)].    \label{det:case4}
    \end{eqnarray}
\end{itemize}
The quantum nonlocal behavior obtained from the inner box after getting mapped to the output behavior with detector imperfection parameters remain nonlocal if the behavior obtained from the outer box violates the inequality~\cite{sauer2020quantum}
\begin{eqnarray} 
&&p(++|00) + p(++|01) + p(++|10) - p(++|11) \nonumber \\
&&- p_A(+|0) - p_B(+|0) \leq 0, \label{SauerInequality}
\end{eqnarray}
where $p_{A}(o|s)$ and $p_{B}(o|s)$ are the probabilities of Alice and Bob to obtain the outcome $o$ on measuring $s$. 

At first, let us assume there is a quantum source in the inner box that is generating a bipartite quantum state whose behavior $\{p(a^{id},b^{id}|xy)\}$ violates the PRBLG MDL inequality given by Eq.~\eqref{MDLeq12} assuming that detectors are perfect. The quantum behavior $\{p(a^{id},b^{id}|xy)\}$ obtained from the inner box gets mapped at the outer box to the behavior $\{p(a^{ob},b^{ob}|xy)\}$ with the introduction of the detector inefficiency $\eta$ and dark count probability $\delta$. The behavior $\{p(a^{ob},b^{ob}|xy)\}$ is then inserted in Eq.~\eqref{SauerInequality} to obtain the critical detector parameters using Algorithm~\ref{algo:Gisin}. For a fixed value of $\delta$ we obtain the minimum value of $\eta$ that violates Eq.~\eqref{SauerInequality} using Algorithm~\ref{algo:Gisin}. We abbreviate left-hand side as LHS.
\begin{algorithm}[H]
\caption{} 
\label{algo:Gisin}
\begin{algorithmic}[1]
\Initialize{$l \gets $ parameter of Eq.~\eqref{MDLeq12}}
\For{$\delta$ in range (0, 1)}
        \For{$\eta$ in range (0, 1)}
            \State $P(a^{id}, b^{id}|xy) \gets $ inner box quantum behavior
            \State $P(a^{ob}, b^{ob}|xy) \gets f(P(a^{id}, b^{id}|xy),\delta,\eta)$ 
            \State obj $\gets$ LHS of Eq.~\eqref{SauerInequality} for $P(a^{ob}, b^{ob}|xy)$
            \State MDL $\gets$ LHS of Eq.~\eqref{MDLeq12} for $P(a^{id}, b^{id}|xy)$
            \State maximize: obj
            \State such that: MDL $> 0$
            \State opt $\gets$ $\max$ value of obj
            \If{opt $> 0$}
                \State print: $\delta$, $\eta$
                \State Break
            \EndIf
        \EndFor
    \EndFor
\end{algorithmic}
\end{algorithm}
We plot the critical values of $\eta$ and $\delta$ obtained using Algorithm~\ref{algo:Gisin} in Fig.~\ref{fig:critParGisin}.
\begin{observation}
From Fig.~\ref{fig:critParGisin}, we see that the minimum value of $\eta$ for a given $\delta$ takes the highest value for $l=0$ and decreases as $l$ increases. For a fixed value of $l$, the minimum value of $\eta$ increases monotonically with the increase in dark count probability. We note that for $\delta = 0$, we have $\eta \approx 0.667$ for all the values of $l$.
\end{observation}

We next assume there is a quantum source in the inner box that is generating a bipartite quantum state whose behavior $\{p(a^{id},b^{id}|xy)\}$ violates the ZRLH MDL inequality given by Eq.~\eqref{RamanathanMDL} assuming that detectors are perfect. The quantum behavior $\{p(a^{id},b^{id}|xy)\}$ obtained from the inner box gets mapped at the outer box to the behavior $\{p(a^{ob},b^{ob}|xy)\}$ with the introduction of the detector inefficiency $\eta$ and dark count probability $\delta$. The behavior $\{p(a^{ob},b^{ob}|xy)\}$ is then inserted in Eq.~\eqref{SauerInequality} to obtain the critical detector parameters using Algorithm~\ref{algo:Ramanathan}. For a fixed value of $\delta$ we obtain the minimum value of $\eta$ that violates Eq.~\eqref{SauerInequality} using Algorithm~\ref{algo:Ramanathan}. We abbreviate left-hand side as LHS.
\begin{algorithm}[H]
\caption{} \label{algo:Ramanathan}
\begin{algorithmic}[1]
\Initialize{$w \gets $ parameter of Eq.~\eqref{RamanathanMDL}\\ 
$l \gets $ parameter of Eq.~\eqref{RamanathanMDL}}
\For{$\delta$ in range (0, 1)}
        \For{$\eta$ in range (0, 1)}
            \State $P(a^{id}, b^{id}|xy) \gets $ inner box quantum behavior
            \State $P(a^{ob}, b^{ob}|xy) \gets f(P(a^{id}, b^{id}|xy),\delta,\eta)$ 
            \State obj $\gets$ LHS of Eq.~\eqref{SauerInequality}
            \State MDL $\gets$ LHS of Eq.~\eqref{RamanathanMDL}
            \State maximize: obj
            \State such that: MDL $> 0$.
            \State opt $\gets$ $\max$ value of obj
            \If{opt $> 0$}
                \State print: $\delta$, $\eta$
                \State Break
            \EndIf
        \EndFor
    \EndFor
\end{algorithmic}
\end{algorithm}
We plot in Fig.~\ref{fig:critParGisin}, the critical values of $\eta$ and $\delta$ from Algorithm~\ref{algo:Ramanathan} for $w=0$.
\begin{observation} \label{obs:3}
The ZRLH MDL inequality given by Eq.~\eqref{RamanathanMDL} and the PRBLG MDL inequality given by Eq.~\eqref{MDLeq12} reduce to the same inequality for $l = 0$ and is independent of $w$. For $l = 0$, Eqs.~\eqref{RamanathanMDL}~and~\eqref{MDLeq12} cannot be violated~\footnote{For $l=0$, Eqs.~\eqref{RamanathanMDL}~and~\eqref{MDLeq12} reduces to the inequality 
    \begin{equation*}
        -[p(+-|01)+p(-+|10)+p(++|11)] \leq 0
    \end{equation*}
    that is independent of $w$. The probabilities are always non-negative, hence the above inequality can never be violated}. For $w = 0$, Eqs.~\eqref{RamanathanMDL}~and~\eqref{MDLeq12} reduce to the same inequality and their violation can happen only when $l > 0$~\footnote{For $w = 0$, violation of Eqs.~\eqref{RamanathanMDL}~and~\eqref{MDLeq12} requires 
    \begin{equation*}
        l > \frac{p(+-|01) + p(-+|10) + p(++|11)}{p(++|00) + 3[p(+-|01) + p(-+|10) + p(++|11)]},
    \end{equation*} i.e., $l > 0$ necessarily.}. For the case $w=0$, we observe same dependence for the threshold detector parameters as can be seen from Fig.~\ref{fig:critParGisin}. We observe from Fig.~\ref{fig:critParGisin} that for a fixed value of $l$, the minimum detection efficiency increases monotonically with the dark count probability.
\end{observation}
We comment here that using Algorithm~\ref{algo:Ramanathan}, one can calculate the detector requirements for other values of $(w, l)$ not mentioned in this section. Next, we consider the state $\rho_g$ and measurement settings $\{A_0^g, A_1^g, B_0^g, B_1^g\}$ with $\theta \approx 1.13557$. This choice of state and measurement settings was shown to ensure $1.6806$ bits of global randomness in \cite{zhao2022tilted}. 
\begin{observation}
Consider the quantum source in the inner box generates bipartite quantum state $\rho_g$. Let Alice and Bob choose the measurement settings $\{A_0^g, A_1^g, B_0^g, B_1^g\}$, with $\theta \approx 1.13557$. With this choice of state and measurement settings we obtain the quantum behavior $\{p(a^{id},b^{id}|xy)\}$ assuming that the detectors are perfect. The output behavior $\{p(a^{ob},b^{ob}|xy)\}$ obtained from the outer box is evaluated using Eq.~\eqref{eq:IdealToObserved}. The behavior $\{p(a^{ob},b^{ob}|xy)\}$ is nonlocal if we have a violation of the inequality
\begin{eqnarray}
&& \delta  \eta  [\delta  (9.50424\delta -4 \delta^2 -9.00848)+4.25636] \nonumber \\
&&+\eta ^2[\delta  (2\delta^3 -5.50424\delta^2 +5.82473\delta-3.13675) \nonumber \\
&& +0.816258]  +2  \delta (\delta -1) (\delta^2 - \delta +1) \nonumber \\
&&-0.752119 \eta \leq 0.
\end{eqnarray}
The allowed values of the pair $(\eta,\delta)$ for which $p(a^{ob},b^{ob}|xy)$ is quantum nonlocal shown in  Fig.~\ref{fig:critParMaxGlobalRandomness}.
\begin{figure}
    \includegraphics[scale=0.6]{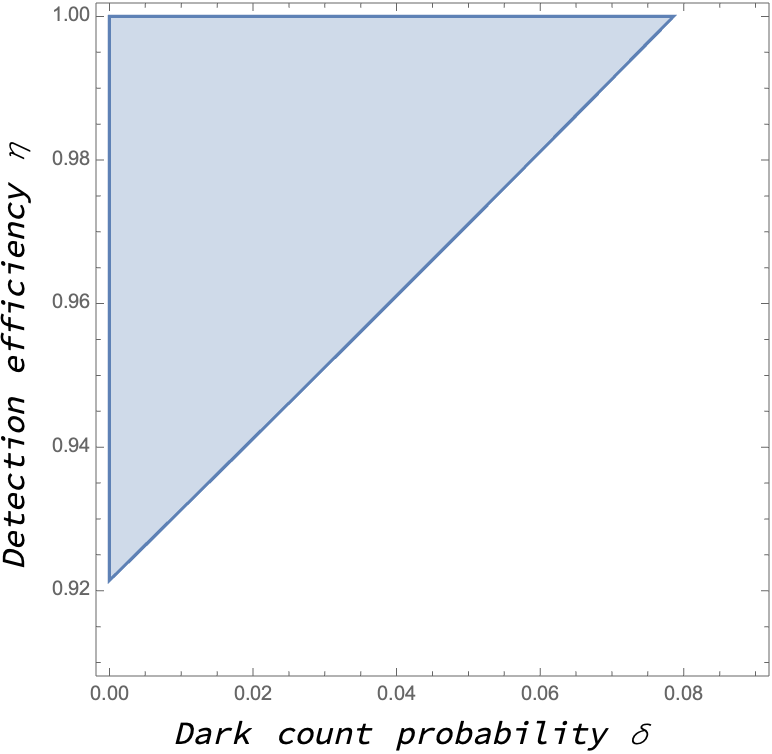}
    \caption{In this figure we plot the detector parameters $\eta$ and $\delta$ for which the behavior $\{p(a^{ob},b^{ob}|xy)\}$ produced in the outer box is quantum nonlocal when the quantum behavior  $\{p(a^{id},b^{id}|xy)\}$ is produced in the inner box using the state $\rho_g$ and measurement settings $\{A_0^g, A_1^g, B_0^g, B_1^g\}$ for $\theta \approx 1.13557$. (Color Online)}
    \label{fig:critParMaxGlobalRandomness}
\end{figure}
In Fig.~\ref{fig:critParMaxGlobalRandomness} we observe that for $\delta = 0$ the minimum detection efficiency $\eta_{\text{crit}} \approx 0.92$ is required to ensure $p(a^{ob},b^{ob}|xy)$ is quantum nonlocal. We also observe that $\eta_{\text{crit}}$ increases monotonically with increase in $\delta$.
\end{observation}

\section{The Tilted Bell inequality} \label{sec:TiltedBell}
The AMP tilted Bell inequality is given as~\cite{acin2012randomness}
\begin{equation} \label{TBI}
    I_{\alpha}^{\beta} := \beta \langle x_1 \rangle + \alpha \langle x_1 y_1 \rangle + \alpha \langle x_1 y_2 \rangle + \langle x_2 y_1 \rangle - \langle x_2 y_2 \rangle \leq \beta + 2\alpha.
\end{equation}
It plays an important role in (a) demonstrating the inequivalence between the amount of certified randomness and the amount of nonlocality~\cite{acin2012randomness}, (b) self-testing of all bipartite pure entangled states~\cite{coladangelo2017all}, (c) protocol for device-independent quantum random number generation with sublinear amount of quantum communication~\cite{bamps2018device}, (d) unbounded randomness certification from a single pair of entangled qubits with sequential measurements~\cite{curchod2017unbounded}. 

In the following proposition, we obtain the bound on $I_{\alpha}^{\beta}$ when the measurement independence assumption is relaxed (see Appendix \ref{TiltedBellBoundProof} for proof).
\begin{proposition}\label{thm:tilt-bound}
The AMP tilted Bell expression $I_{\alpha}^{\beta}$ in the presence of locality and the relaxed measurement independence is bounded by 
\begin{equation} 
    I_{\alpha}^{\beta} \leq \beta + 2 \alpha + \min\{\alpha (M_1 + \min\{M_1,M_2\}) + M_2, 2\}, \label{TBILUpperBound}
\end{equation}
where $M_1$ and $M_2$ are the measurement dependence parameters for Alice and Bob. 
\end{proposition}
\subsection{Comparison of one-sided and two-sided measurement dependence to ensure quantum representation}
Let Alice and Bob have free will in choosing the measurement settings, then the maximum violation of $I_{\alpha}^{\beta}$ obtained by quantum nonlocal behaviors is given by~\cite{acin2012randomness}
\begin{equation} \label{check}
    I_{\alpha}^{\beta} \leq 2 \sqrt{(1 + \alpha^2)(1 + \frac{\beta^2}{4})}.
\end{equation}
As direct consequences of Proposition~\ref{thm:tilt-bound}, we have following corollaries.
\begin{corollary}
When $M_1$ = $M_2$ = $M$, the quantum nonlocal behaviors that maximally violate Eq.~\eqref{TBI} with the amount of violation given by the RHS of Eq.~\eqref{check}, remains nonlocal for 
\begin{equation} \label{secVcor1}
    M < \frac{- 2\alpha - \beta + \sqrt{(1 + \alpha^2)(4 + \beta^2)}}{1 + 2\alpha}.
\end{equation}
\end{corollary}
For $\alpha = 1 \text{ and } \beta = 0$ (the Bell-CHSH inequality), Eq.~\eqref{secVcor1} reduces to $M < \frac{2}{3}(\sqrt{2}-1) \approx 0.276$ and is consistent with the observation in~\cite{friedman2019relaxed}. 
\begin{corollary}
When $M_1 = 0$ and $M_2 = M$, the quantum nonlocal behaviors that maximally violate Eq.~\eqref{TBI} with the amount of violation given by the RHS of Eq.~\eqref{check}, remains nonlocal for 
\begin{equation} \label{secVcor2}
    M < - 2\alpha - \beta + \sqrt{(1 + \alpha^2)(4 + \beta^2)}.
\end{equation}
\end{corollary}
For $\beta = 0 \text{ and } \alpha = 1$ (the Bell-CHSH inequality), Eq.~\eqref{secVcor2} reduces to $M < 2(\sqrt{2}-1) \approx 0.828$ and is consistent with the observation in \cite{friedman2019relaxed}. 
\begin{corollary}
When $M_1 = M$ and $M_2 = 0$, the quantum nonlocal behaviors that maximally violate Eq.~\eqref{TBI} with the amount of violation given by the RHS of  Eq.~\eqref{check}, remains nonlocal for 
\begin{equation} \label{secVcor3}
    M < \frac{1}{\alpha}(- 2\alpha - \beta + \sqrt{(1 + \alpha^2)(4 + \beta^2)}).
\end{equation}
\end{corollary} 
For $\beta = 0 \text{ and } \alpha = 1$ (the Bell-CHSH inequality), Eq.~\eqref{secVcor3} reduces to $M < 2(\sqrt{2}-1) \approx 0.828$ and is consistent with the observation in \cite{friedman2019relaxed}. 

\begin{figure}
    \includegraphics[scale = 0.9]{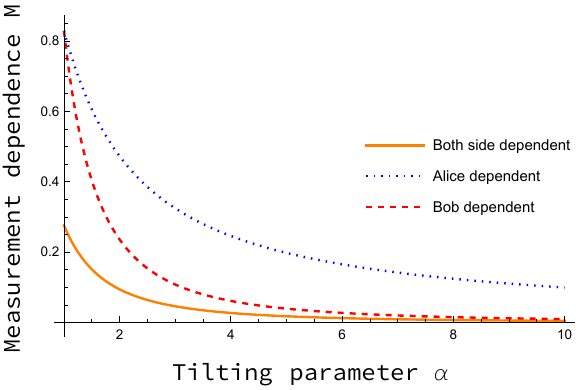}
    \caption{In this figure, we consider $\beta = 0$ and plot the maximum values of the measurement dependence parameter $M$ as a function of the tilting parameter $\alpha$ in Eq.\eqref{TBI}. The values of $M$ and $\alpha$ in this figure ensures that the quantum nonlocal behaviors that maximally violate Eq.~\eqref{TBI} remains nonlocal for the cases of (a) both-sided measurement dependence (in orange) (b) only Alice has measurement dependence (in dotted blue line) (c) only Bob has measurement dependence (in dashed red line.) (Color Online)} 
    \label{fig:critValueMAlpha}
\end{figure}

In Fig.~\ref{fig:critValueMAlpha}, we plot an upper bound on $M$ as a function of $\alpha$ for MDL violating behaviors when $\beta = 0$ and $\alpha \geq 1$. The values of $M$ and $\alpha$ from Fig.~\ref{fig:critValueMAlpha} ensures that the quantum nonlocal behaviors that maximally violate Eq.~\eqref{TBI} remains nonlocal in the presence of measurement dependence.
We observe in Fig.~\ref{fig:critValueMAlpha} that for $\beta = 0$ and $\alpha \geq 1$, the introduction of one-sided measurement dependence allows for higher values of $M$ (implying lower degree of freedom-of-choice) as compared to introducing both-sided measurement dependence. Also, for the case of one-sided measurement dependence, Alice can have higher values of $M$ as compared to Bob. 
\subsection{Bounds on the measurement dependence for testing nonlocality}
We observe in Proposition~\ref{thm:tilt-bound} that in the presence of relaxed measurement dependence, the behaviors $\{p(ab|xy)\}$, that violates Eq.~\eqref{NL} are nonlocal.
\begin{equation} \label{NL}
    I_{\alpha}^{\beta} \leq \beta + 2 \alpha + \min\{\alpha(M_1 + \min\{M_1, M_2\}) + M_2, 2\}
\end{equation}
In the following, we discuss some of the cases where Eq.~\eqref{NL} can be violated.
\begin{itemize}
    \item[a.] We consider a situation when both Alice and Bob have the same measurement dependence, we have $M_1 = M_2 = M$. For this case, violating Eq.~\eqref{NL} requires
\begin{equation} \label{NL1}
    I_{\alpha}^{\beta} > \beta + 2 \alpha + \min\{(2 \alpha + 1) M, 2\}.
\end{equation}
If $(2 \alpha + 1) M \geq 2$, $I_{\alpha}^{\beta}$ reaches the no-signalling boundary. Whereas, if $(2 \alpha + 1) M < 2$, then Eq.~\eqref{NL1} reduces to inequality 
\begin{equation}
    M < \frac{I_{\alpha}^{\beta} - \beta - 2 \alpha}{2 \alpha + 1}.
\end{equation}
\item[b.] We consider the situation when only Bob has measurement dependence, we have $M_1 = 0, M_2 = M$. For this case, violating Eq.~\eqref{NL} requires
    \begin{equation} \label{C51}
        I_{\alpha}^{\beta} > \beta + 2 \alpha + \min\{M, 2\}.
    \end{equation}
    If $ M = 2$, $I_{\alpha}^{\beta}$ reaches the no-signalling boundary. Whereas, if $M < 2$, then Eq.~\eqref{C51} to the inequality 
    \begin{equation}
        M < I_{\alpha}^{\beta} - \beta - 2 \alpha.
    \end{equation}
\item[c.] We consider the situation when only Alice has measurement dependence, we have $M_1 = M, M_2 = 0$. For this case, violating Eq.~\eqref{NL} requires
    \begin{equation} \label{C61}
        I_{\alpha}^{\beta} > \beta + 2 \alpha + \min\{\alpha M, 2\}.
    \end{equation}
    If $\alpha M \geq 2$, $I_{\alpha}^{\beta}$ reaches the no-signalling boundary. Whereas if $\alpha M < 2$, then Eq.~\eqref{C61} reduces to the inequality 
    \begin{equation}
        M < \frac{I_{\alpha}^{\beta} - \beta - 2 \alpha}{\alpha }.
    \end{equation}
\end{itemize}
We note that for $\alpha = 1$ and $\beta = 0$, i.e., for the Bell-CHSH inequality, the values of $M_1$ and $M_2$ that violates Eq.~\eqref{NL} are in agreement with that obtained in \cite{friedman2019relaxed}. 

\begin{figure}
    \includegraphics[scale = 0.6]{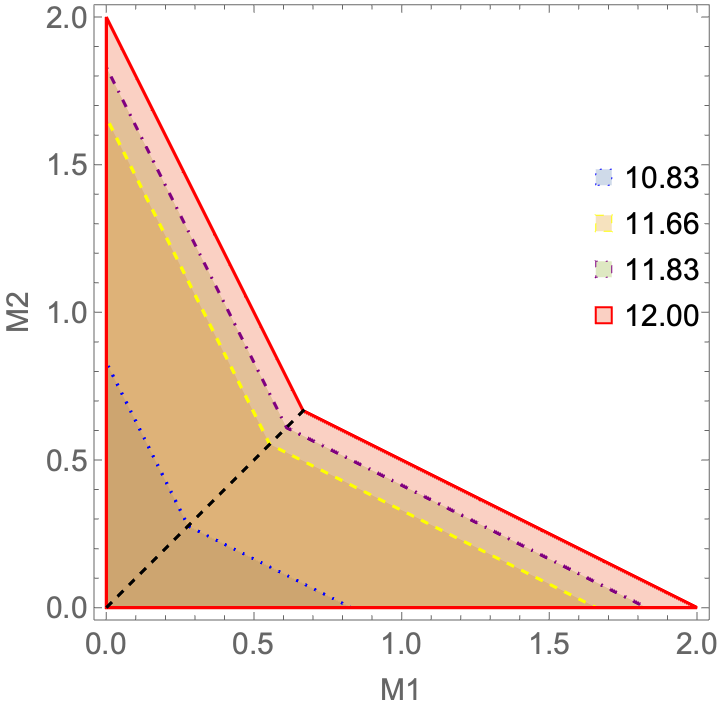}
    \caption{In this figure, we plot the values of the measurement dependence parameters $M_1$ and $M_2$ for which the violation of Eq.~\eqref{TBI} given by $I_{\alpha}^{\beta} = I'(\alpha = 1,\beta = 8,\textbf{P}') \in \{10.83, 11.66, 11.83, 12.00\}$ cannot be described by a deterministic MDL model. The correlations violating Eq.~\eqref{TBI} with (a) $I_{\alpha}^{\beta} = 12.00$ belong to the no-signalling boundary (shown enclosed by red line) (b) $I_{\alpha}^{\beta} = 11.66$ belong to the quantum boundary (shown enclosed by dashed yellow line) (c) $I_{\alpha}^{\beta} = 10.83$ belong to the quantum set (shown enclosed by dotted blue line), and (d) $I_{\alpha}^{\beta} = 11.83$ belong to the no-signalling set (shown enclosed by dot-dashed purple line). The black line in the figure denotes equal values of $M_1$ and $M_2$ in the regions (a), (b), (c), and (d). (Color online)}
    \label{fig:M1M2TiltedBell}
\end{figure}

Consider there exists some behavior $\mathbf{P}'=\{p'(ab|xy)\}$ that violates Eq.~\eqref{TBI} with the amount of violation given by $I_{\alpha}^{\beta}=I'(\alpha,\beta,\mathbf{P}')$. For $\alpha = 1,\beta = 8$, we show in Fig.~\ref{fig:M1M2TiltedBell} the possible values of measurement dependence parameters $M_1$ and $M_2$ for which the amount of violation of Eq.~\eqref{TBI} given by $I'$ cannot be described by a deterministic measurement dependent local model. The plots for the other combinations of $(\alpha,\beta)$ is given in Appendix \ref{DIBounds}. 

\section{Discussion}
In this paper, two different approaches in quantifying measurement dependence via parameters $l$ and $(M_1,M_2)$ from standard literature is presented and a bound on these parameters for certifying nonlocality of different behaviors is obtained. It is observed that the quantum behavior certifying close to 2 bits of randomness remains measurement dependent nonlocal only in the limit of complete measurement independence, i.e., in the limit of $l \rightarrow 0.25$. The quantum behavior that provides 1.6806 bits of global randomness from the violation of the tilted Hardy relations is measurement dependent nonlocal for arbitrarily small values of $l$. This motivates further study on obtaining an inequality that provides close to two bits of global randomness in the limit of arbitrarily low measurement dependence, a problem we leave for future study. Deviating from the conventional approach of assuming perfect detectors, we present a framework to determine the threshold values of the detector parameters that are robust enough to certify nonlocality of given quantum behaviors. This is an important step towards experimentally obtaining nonlocality in the presence of relaxed measurement independence. For an illustration, we presented the critical requirements for generating 1.6806 bits of tilted-Hardy certified global randomness. The detector parameters obtained from this study is expected to have important applications in experimentally implementing various information processing tasks that rely on quantum nonlocality as a resource.

The modified analytical bound on the AMP tilted Bell inequality in terms of $M_1$ and $M_2$ has been obtained. Using the analytical bound, the bounds on $M_1$ and $M_2$ to ensure quantumness and nonlocality are observed. It is observe that one-sided measurement dependence is more advantageous from the point of view of the user as compared to two-sided measurement dependence. The analytical bound obtained is expected to have applications in self-testing of quantum states and other device independent information processing protocols like randomness generation and secure communication.

For future work, it would be interesting to see implications of measurement dependence in multipartite Bell-type inequalities~\cite{brunner2014bell,HSD15} for the certification of multipartite nonlocality and device-independent conference keys~\cite{RMW18,holz2020genuine,HWD22}. 

\textit{Note}--- We noticed the related work ``Quantum nonlocality in presence of strong measurement dependence" by Ivan Šupić, Jean-Daniel Bancal, and Nicolas Brunner recently posted as arXiv:2209.02337.
\begin{acknowledgements}
The authors thank Karol Horodecki for discussions. The authors thank the anonymous Referee for pointing out a bug in the previous version of Fig.~\ref{fig:critParGisin} and Observation~\ref{obs:3}, which now are corrected. A.S. acknowledges the PhD fellowship from the Raman Research Institute, Bangalore. A.S. thanks the International Institute of Information Technology, Hyderabad for the hospitality during his visit in the summer of 2022 where part of this work was done.
\end{acknowledgements}
\begin{appendix}
\section{Calculating the modified local bound for the AMP tilted Bell inequality}\label{TiltedBellBoundProof}
We consider the AMP tilted-Bell inequality introduced in \cite{acin2012randomness} and given by,
\begin{equation} \label{Tb}
    I_{\alpha}^{\beta} = \beta \langle x_1 \rangle + \alpha \langle x_1 y_1 \rangle + \alpha \langle x_1 y_2 \rangle + \langle x_2 y_1 \rangle - \langle x_2 y_2 \rangle
\end{equation}
The determinism assumption states that the measurement outcomes are deterministic functions of the choice of settings and the hidden variable $\lambda$, i.e., $a = A(x,\lambda)$ and $b = B(x,\lambda)$ with $p(a|x\lambda) = \delta_{a,A(x,\lambda)}$ and $p(b|y\lambda) = \delta_{b,B(y,\lambda)}$. If we assume that the choice of the measurement settings on the side of Alice and Bob can depend on some hidden variable, $\lambda$, the correlations of Alice and Bob can be expressed as
\begin{equation} \label{corrAB}
    \langle x  y \rangle = \int {\rm d}\lambda p(\lambda|x,y)  A(x,\lambda)  B(y,\lambda).
\end{equation}
The correlation function of Alice can also be written as
\begin{equation}\label{corrA}
    \langle x \rangle = \int {\rm d}\lambda  p(\lambda|x)  A(x,\lambda). 
\end{equation}
Applying Eq.~\eqref{corrAB} and Eq.~\eqref{corrA} to Eq.~\eqref{Tb} we have
\begin{eqnarray} \label{Tb1}
I_{\alpha}^{\beta} = \beta \int && {\rm d}\lambda  p(\lambda|x_1)  A(x_1,\lambda) \nonumber \\
&& + \alpha \int {\rm d}\lambda  p(\lambda|x_1,y_1)  A(x_1,\lambda)  B(y_1,\lambda) \nonumber \\
&& + \alpha \int {\rm d}\lambda  p(\lambda|x_1,y_2)  A(x_1,\lambda)  B(y_2,\lambda)  \nonumber \\
&&+ \int {\rm d}\lambda  p(\lambda|x_2,y_1)  A(x_2,\lambda)  B(y_1,\lambda) \nonumber \\
&& -\int {\rm d}\lambda  p(\lambda|x_2,y_2)  A(x_2,\lambda)  B(y_2,\lambda).
\end{eqnarray}
Adding and subtracting the terms 
\begin{eqnarray}
&& \alpha \int {\rm d}\lambda  p(\lambda|x_1,y_2)  A(x_1,\lambda)  B(y_1,\lambda) \text{ and} \nonumber\\
&&\int {\rm d}\lambda  p(\lambda|x_2,y_2)  
A(x_2,\lambda)  B(y_1,\lambda) \nonumber  
\end{eqnarray}
to Eq.~\eqref{Tb1} we obtain
\begin{eqnarray} \label{tb4}
I_{\alpha}^{\beta} &&= \beta \int {\rm d}\lambda  p(\lambda|x_1)  A(x_1,\lambda) \nonumber \\ 
&& \qquad + \alpha \int {\rm d}\lambda  A(x_1,\lambda)  B(y_1,\lambda)  \bigg[p(\lambda|x_1,y_1) \nonumber \\
&& \qquad \qquad \qquad \qquad \qquad \qquad - p(\lambda|x_1,y_2)\bigg] \nonumber\\
&& \qquad + \int {\rm d}\lambda  A(x_2,\lambda)  B(y_1,\lambda)  \bigg[p(\lambda|x_2,y_1) \nonumber \\
&& \qquad \qquad \qquad \qquad \qquad \qquad - p(\lambda|x_2,y_2)\bigg] \nonumber \\
&& \qquad + \alpha \int {\rm d}\lambda  p(\lambda|x_1,y_2)  \bigg[A(x_1,\lambda)  B(y_2,\lambda) \nonumber \\
&& \qquad \qquad \qquad \qquad \qquad + A(x_1,\lambda)  B(y_1,\lambda)\bigg] \nonumber \\
&& \qquad -\int {\rm d}\lambda  p(\lambda|x_2,y_2)  \bigg[A(x_2,\lambda)  B(y_2,\lambda) \nonumber \\
&& \qquad \qquad \qquad \qquad - A(x_2,\lambda)  B(y_1,\lambda)\bigg]. 
\end{eqnarray}
We observe that Eq.~\eqref{tb4} is bounded by $I_{\alpha}^{\beta} \leq \max[T_1] + \max[T_2] + \max[T_3]$. We also note that as $p(\lambda|x_1)$ is a normalised probability distribution, $\int {\rm d}\lambda p(\lambda|x_1) = 1$. This implies that the maximum value of the quantity $\beta \int {\rm d}\lambda  p(\lambda|x_1)  A(x_1,\lambda)$ is given by $\beta$ when $A(x_1,\lambda)$ is set to $1$. With these observations, $T_1$ can be simplified as
\begin{eqnarray}
T_1 = &&\beta \int {\rm d}\lambda  p(\lambda|x_1)  A(x_1,\lambda) \nonumber \\
&&+ \alpha \int {\rm d}\lambda  A(x_1,\lambda)  B(y_1,\lambda)  \bigg[p(\lambda|x_1,y_1) \nonumber\\
&&\qquad\qquad - p(\lambda|x_1,y_2)\bigg] \leq \beta + \alpha M_2. \label{eq:T1bound}
\end{eqnarray}
where we have set $B(y_1,\lambda) = 1$. To evaluate the maximum value of $T_2$ we set $A(x_2,\lambda) = 1$ and obtain 
\begin{align}
T_2 = & \int {\rm d}\lambda  A(x_2,\lambda) B(y_1,\lambda)\bigg[p(\lambda|x_2,y_1) - p(\lambda|x_2,y_2)\bigg]\nonumber \\
    & \leq M_2. \label{eq:T2bound}
\end{align}
We evaluate $T_3$ as follows, 
\begin{eqnarray}
&&T_3 \nonumber \\
 = && \alpha \int {\rm d}\lambda p(\lambda|x_1,y_2) \bigg[A(x_1,\lambda) B(y_2,\lambda) + A(x_1,\lambda) B(y_1,\lambda)\bigg] \nonumber \\
&& -\int {\rm d}\lambda p(\lambda|x_2,y_2) \bigg[A(x_2,\lambda) B(y_2,\lambda) - A(x_2,\lambda) B(y_1,\lambda)\bigg] \nonumber \\ 
 =&& \int {\rm d}\lambda A(x_1,\lambda) B(y_2,\lambda) \bigg[\alpha p( \lambda|x_1,y_2) \nonumber \\
&& \qquad \qquad\qquad \qquad - p(\lambda|x_2,y_2) \frac{A(x_2,\lambda)}{A(x_1,\lambda)}\bigg] \nonumber \\
&&+ \int {\rm d}\lambda A(x_1,\lambda) B(y_1,\lambda) \bigg[\alpha p( \lambda|x_1,y_2) \nonumber \\
&& \qquad \qquad\qquad \qquad + p(\lambda|x_2,y_2) \frac{A(x_2,\lambda)}{A(x_1,\lambda)}\bigg].
\end{eqnarray}
To get the maximum value of $T_3$, we set the values of $A(x_1,\lambda), B(y_1,\lambda), A(x_1,\lambda), B(y_2,\lambda)$  to one. This then implies, 
\begin{eqnarray}
T_3 = &&  \int {\rm d}\lambda \bigg[\alpha  p(\lambda|x_1,y_2) - p(\lambda|x_2,y_2)\bigg] \nonumber \\
&& + \int {\rm d}\lambda  \bigg[\alpha  p(\lambda|x_1,y_2) + p(\lambda|x_2,y_2)\bigg]. \label{T3}
\end{eqnarray}
We evaluate the first term of the Eq.~\eqref{T3} as
\begin{eqnarray}
&&\int {\rm d}\lambda  \bigg[\alpha  p(\lambda|x_1,y_2) - p(\lambda|x_2,y_2)\bigg] \nonumber \\
=&&  \int {\rm d}\lambda  \bigg[\alpha  p(\lambda|x_1,y_2) - \alpha  p(\lambda|x_2,y_2) + \alpha  p(\lambda|x_2,y_2) \nonumber \\
&& \qquad - p(\lambda|x_2,y_2)\bigg]  \\ 
=&&  \int {\rm d}\lambda  \bigg[\alpha  \bigg(p(\lambda|x_1,y_2) -  p(\lambda|x_2,y_2)\bigg) \nonumber \\
&& \qquad + (\alpha - 1)  p(\lambda|x_2,y_2)\bigg]  \\
=&&  \alpha \int {\rm d}\lambda  \bigg[p(\lambda|x_1,y_2) -  p(\lambda|x_2,y_2)\bigg] \nonumber \\
&& \qquad + (\alpha - 1) \int {\rm d}\lambda   p(\lambda|x_2,y_2). \label{app:T1} 
\end{eqnarray}
In Eq.~\eqref{app:T1} we note that
\begin{equation*}
    \alpha \int {\rm d}\lambda \bigg(p(\lambda|x_1,y_2) -  p(\lambda|x_2,y_2)\bigg) \leq \alpha M_1.
\end{equation*} 
We note that as $p(\lambda|x_2,y_2)$ is a normalised probability distribution, $\int {\rm d}\lambda p(\lambda|x_2,y_2) = 1$. This then implies
\begin{equation}
    \int {\rm d}\lambda  \bigg[\alpha  p(\lambda|x_1,y_2) - p(\lambda|x_2,y_2)\bigg]  \leq  \alpha  M_1 +  \alpha  -  1,
\end{equation}
which on simplification reduces to
\begin{equation} \label{T31}
    \int {\rm d}\lambda  \bigg[\alpha  p(\lambda|x_1,y_2) - p(\lambda|x_2,y_2) \bigg]  \leq  \alpha  (M_1 + 1) -  1.
\end{equation}
We proceed in the same way for the second term in Eq.~\eqref{T3} as follows,
\begin{eqnarray}
&&\int {\rm d}\lambda  \bigg[\alpha  p(\lambda|x_1,y_2) + p(\lambda|x_2,y_2)\bigg] \nonumber \\
=&& \int {\rm d}\lambda  \bigg[\alpha  p(\lambda|x_1,y_2) + \alpha  p(\lambda|x_2,y_2) - \alpha  p(\lambda|x_2,y_2) \nonumber \\
&& \qquad + p(\lambda|x_2,y_2)\bigg]  \\
=&& \int {\rm d}\lambda  \alpha  \bigg[p(\lambda|x_1,y_2) +  p(\lambda|x_2,y_2)\bigg] \nonumber \\
&& \qquad - (\alpha - 1) \int {\rm d}\lambda p(\lambda|x_2,y_2) \\
=&& \alpha \int {\rm d}\lambda  \bigg[p(\lambda|x_1,y_2) +  p(\lambda|x_2,y_2)\bigg] \nonumber \\
&& \qquad + (1 - \alpha) \int {\rm d}\lambda   p(\lambda|x_2,y_2). 
\end{eqnarray}
We note that as $p(\lambda|x_1,y_2)$, $p(\lambda|x_2,y_2)$ are normalised probability distributions, $\int {\rm d}\lambda p(\lambda|x_1,y_2) = 1$ and $\int {\rm d}\lambda p(\lambda|x_2,y_2) = 1$. This then implies
\begin{equation} \label{T32}
    \int {\rm d}\lambda  \bigg[\alpha  p(\lambda|x_1,y_2) + p(\lambda|x_2,y_2) \bigg]  =  \alpha + 1.
\end{equation}
If we insert Eq.~\eqref{T31}~and~Eq.~\eqref{T32} in Eq.~\eqref{T3}, we obtain
\begin{eqnarray}\label{T33}
T_3  \leq  \alpha  (M_1 + 2).
\end{eqnarray}
Now combining Eq.~\eqref{eq:T1bound},~Eq.~\eqref{eq:T2bound}~and~Eq.~\eqref{T33}, we have the bound on $I_{\alpha}^{\beta}$ as
\begin{eqnarray} 
I_{\alpha}^{\beta}  && \leq  T_1 + T_2 + T_3  \\
&& \leq \beta + \alpha M_2 + M_2 + \alpha (M_1 + 2)  \\
&& \leq \beta + 2 \alpha +  \alpha M_1 + (\alpha + 1) M_2  \label{Tbf1}  \\
&& \leq \beta + 2 \alpha + \alpha (M_1 + M_2) + M_2. \label{Eq1}
\end{eqnarray}
We again start by considering Eq.~\eqref{Tb1} as 
\begin{eqnarray} 
I_{\alpha}^{\beta} = &&  \beta \int {\rm d}\lambda  p(\lambda|x_1)  A(x_1,\lambda) \nonumber \\
&& \qquad +  \alpha \int {\rm d}\lambda  p(\lambda|x_1,y_1)  A(x_1,\lambda)  B(y_1,\lambda) \nonumber \\
&& \qquad + \alpha \int {\rm d}\lambda  p(\lambda|x_1,y_2)  A(x_1,\lambda)  B(y_2,\lambda)  \nonumber \\
&& \qquad + \int {\rm d}\lambda  p(\lambda|x_2,y_1)  A(x_2,\lambda)  B(y_1,\lambda) \nonumber \\
&& \qquad - \int {\rm d}\lambda  p(\lambda|x_2,y_2)  A(x_2,\lambda)  B(y_2,\lambda).
\end{eqnarray}
Adding and subtracting the terms $\alpha \int {\rm d}\lambda  p(\lambda|x_2,y_1)  A(x_1,\lambda)  B(y_1,\lambda)$ and $\alpha \int {\rm d}\lambda  p(\lambda|x_2,y_2)  A(x_1,\lambda)  B(y_2,\lambda)$ to Eq.~\eqref{Tb1} we obtain
\begin{eqnarray} \label{Tb4}
I_{\alpha}^{\beta} &&= \beta \int {\rm d}\lambda  p(\lambda|x_1)  A(x_1,\lambda) \nonumber \\
&& \qquad + \alpha \int {\rm d}\lambda  A(x_1,\lambda)  B(y_1,\lambda)  \bigg[p(\lambda|x_1,y_1) \nonumber \\
&& \qquad \qquad \qquad \qquad \qquad- p(\lambda|x_2,y_1)\bigg] \nonumber \\
&& \qquad + \alpha \int {\rm d}\lambda  A(x_1,\lambda)  B(y_2,\lambda)  \bigg[p(\lambda|x_1,y_2) \nonumber \\
&&\qquad \qquad \qquad \qquad \qquad - p(\lambda|x_2,y_2)\bigg] \nonumber \\
&& \qquad + \int {\rm d}\lambda  A(x_2,\lambda)  B(y_1,\lambda)\bigg[p(\lambda|x_2,y_1)  \nonumber \\
&& \qquad \qquad \qquad - \frac{B(y_2,\lambda)}{B(y_1,\lambda)}  p(\lambda|x_2,y_2)\bigg] \nonumber \\
&& \qquad + \alpha \int {\rm d}\lambda  A(x_1,\lambda)  B(y_1,\lambda) \bigg[p(\lambda|x_2,y_1)  \nonumber \\
&& \qquad \qquad \qquad + \frac{B(y_2,\lambda)}{B(y_1,\lambda)}  p(\lambda|x_2,y_2)\bigg].
\end{eqnarray}
We observe that Eq.~\eqref{Tb4} is bounded by $I_{\alpha}^{\beta} \leq \max[t_1] + \max[t_2] + \max[t_3]$. We also note that as $p(\lambda|x_1)$ is a normalised probability distribution, $\int {\rm d}\lambda p(\lambda|x_1) = 1$. This implies that the maximum value of the quantity $\beta \int {\rm d}\lambda  p(\lambda|x_1)  A(x_1,\lambda)$ then takes the value of $\beta$ when $A(x_1,\lambda)$ is set to $1$. With these observations, $t_1$ can be simplified as, 
\begin{eqnarray}
t_1 &&= \beta \int {\rm d}\lambda  p(\lambda|x_1)  A(x_1,\lambda) \nonumber \\
&& \qquad + \alpha \int {\rm d}\lambda  A(x_1,\lambda)  B(y_1,\lambda) \bigg[p(\lambda|x_1,y_1) \nonumber \\
&& \qquad \qquad \qquad \qquad \qquad \qquad \qquad - p(\lambda|x_2,y_1)\bigg] \nonumber \\
&&\leq \beta + \alpha M_1.
\end{eqnarray}
Similarly, $t_2$ can be simplified as 
\begin{eqnarray}
t_2 &&= \alpha \int {\rm d}\lambda  A(x_1,\lambda)  B(y_2,\lambda) \bigg[p(\lambda|x_1,y_2) \nonumber \\ 
&&\qquad \qquad \qquad \qquad \qquad \qquad - p(\lambda|x_2,y_2)\bigg] \nonumber \\
&&\leq \alpha M_1,
\end{eqnarray}
and for $t_3$, we have the expression
\begin{eqnarray}
t_3 && = \int {\rm d}\lambda  A(x_2,\lambda) B(y_1,\lambda) \bigg[p(\lambda|x_2,y_1) \nonumber\\ 
&& \qquad \qquad \qquad - \frac{B(y_2,\lambda)}{B(y_1,\lambda)}  p(\lambda|x_2,y_2)\bigg] \nonumber \\
&&\qquad + \alpha \int {\rm d}\lambda  A(x_1,\lambda)  B(y_1,\lambda) \bigg[p(\lambda|x_2,y_1) \nonumber \\ 
&& \qquad \qquad \qquad + \frac{B(y_2,\lambda)}{B(y_1,\lambda)} p(\lambda|x_2,y_2)\bigg]. 
\end{eqnarray}
We set the values of $A(x_1,\lambda), B(y_1,\lambda), A(x_2,\lambda), B(y_2,\lambda)$ to one and obtain
\begin{eqnarray}
t_3 &&=  \int {\rm d}\lambda  \bigg[p(\lambda|x_2,y_1) - p(\lambda|x_2,y_2)\bigg] \nonumber \\
&& \qquad + \alpha \int {\rm d}\lambda \bigg[p(\lambda|x_2,y_1) + p(\lambda|x_2,y_2)\bigg]. \label{t3}
\end{eqnarray}
The first term in Eq,~\eqref{t3} is bounded by
\begin{equation}
 \int {\rm d}\lambda \bigg[p(\lambda|x_2,y_1) - p(\lambda|x_2,y_2)\bigg] \leq M_2.   
\end{equation}
We note that as $p(\lambda|x_2,y_1)$ and $p(\lambda|x_2,y_2)$ are normalised probability distribution, $\int {\rm d}\lambda p(\lambda|x_2,y_1) = 1$ and $\int {\rm d}\lambda p(\lambda|x_2,y_2) = 1$. Eq.~\eqref{t3} is then expressed as
\begin{equation}
    t_3 \leq M_2 + 2 \alpha.
\end{equation}
Now combining the values of $t_1$, $t_2$ and $t_3$, we have the bound on $I_{\alpha}^{\beta}$ as
\begin{eqnarray} 
I_{\alpha}^{\beta} && \leq t_1 + t_2 + t_3  \\
&& \leq \beta + \alpha M_1 + \alpha M_1 + M_2 + 2 \alpha  \\
&& \leq \beta + 2 \alpha +  2 \alpha M_1 + M_2.  \label{Eq2}
\end{eqnarray}
We note that the bound on $I_{\alpha}^{\beta}$ must be the minimum of Eqs.~\eqref{Eq1}~and~\eqref{Eq2} and is expressed as
\begin{equation} \label{tiltedBound}
    I_{\alpha}^{\beta} \leq \beta + 2 \alpha + \alpha [M_1 + \min\{M_1,M_2\}] + M_2.
\end{equation}
We observe that for the case of $\beta = 0$ and $\alpha = 1$,
\begin{equation}
    I_{1}^{0} \leq 2 +  M_1 + M_2 + \min\{M_1, M_2\}, 
\end{equation}
which is in agreement with that obtained in \cite{friedman2019relaxed}. We note that the maximum value that $I_{\alpha}^{\beta}$ [Eq.~\eqref{Tb}] can take is $\beta + 2 \alpha + 2$ (We get this bound by setting $\langle x_1 \rangle = 1$, $\langle x_1 y_1 \rangle = 1$, $\langle x_1 y_2 \rangle = 1$, $\langle x_2 y_1 \rangle = 1$, $\langle x_2 y_2 \rangle = -1$) and arrive at 
\begin{equation} \label{TBNS}
    I_{\alpha}^{\beta} \leq \beta + 2 \alpha + \min\{\alpha (M_1 + \min\{M_1,M_2\}) + M_2, 2\}.
\end{equation}

\section{Bounds on the measurement dependence to certify nonlocality}\label{DIBounds}
\begin{figure}[h]
    \includegraphics[scale = 0.5]{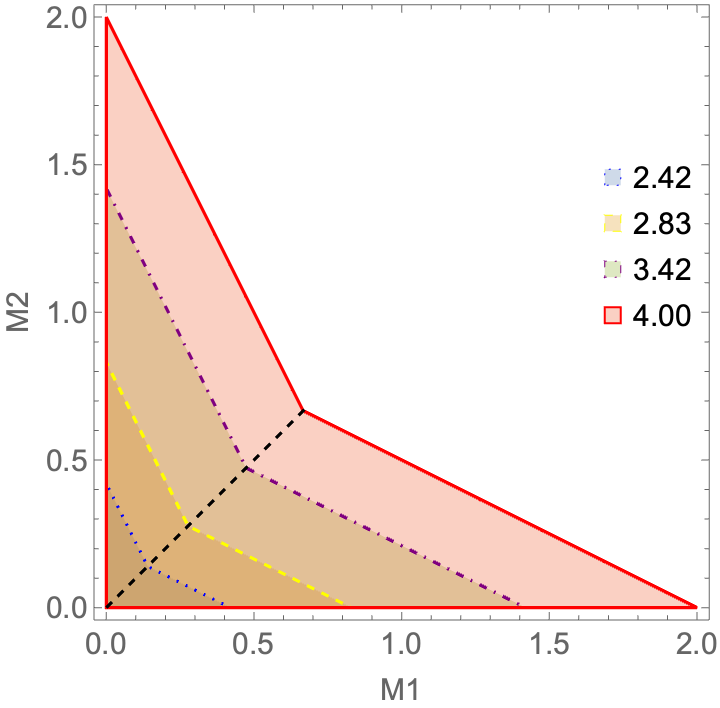}
    \caption{In this figure, we plot the values of the measurement dependence parameters $M_1$ and $M_2$ for which the violation of Eq.~\eqref{TBI} given by $I_{\alpha}^{\beta} = I'(\alpha = 1,\beta = 0,\textbf{P}') \in \{4.00, 3.42, 2.83, 2.42\}$ cannot be described by a deterministic MDL model. The correlations violating Eq.~\eqref{TBI} with (a) $I_{\alpha}^{\beta} = 4.00$ belong to the no-signalling boundary (shown enclosed by red) (b) $I_{\alpha}^{\beta} = 2.83$ belong to the quantum boundary (shown enclosed by dashed yellow line) (c) $I_{\alpha}^{\beta} = 2.42$ belong to the quantum set (shown enclosed by blue dotted), and (d) $I_{\alpha}^{\beta} = 3.42$ belong to the no-signalling set (shown enclosed by dot-dashed purple line). The black line in the figure denotes equal values of $M_1$ and $M_2$ in the regions (a), (b), (c), and (d). (Color Online)}
    \label{appc:fig1}
\end{figure}
We observe in Proposition~\ref{thm:tilt-bound} that in the presence of relaxed measurement dependence, the behaviors $\{p(ab|xy)\}$, that violates Eq.~\eqref{app:NL} are nonlocal.
\begin{equation} \label{app:NL}
    I_{\alpha}^{\beta} \leq \beta + 2 \alpha + \min\{\alpha(M_1 + \min\{M_1, M_2\}) + M_2, 2\}
\end{equation}
Consider there exists some behavior $\mathbf{P}'=\{p'(ab|xy)\}$ that violates Eq.~\eqref{TBI} with the amount of violation given by $I_{\alpha}^{\beta}=I'(\alpha,\beta,\mathbf{P}')$. We show in Fig.~\ref{appc:fig1} the possible values of measurement dependence parameters $M_1$ and $M_2$ for which the amount of violation of Eq.~\eqref{TBI} given by $I'$ cannot be described by a deterministic measurement dependent local model. We observe that Fig.~\eqref{appc:fig1} is in agreement with that obtained in \cite{friedman2019relaxed}. 
\end{appendix}

\bibliographystyle{unsrt}
\bibliography{refs}{}
\end{document}